\documentclass{ws-ijbc}
\usepackage{ws-rotating}     
\usepackage{graphicx}
\usepackage{epstopdf}

\def\l{\lambda}
\def\al{\alpha}

\def\ie{i.e.}
\def\eg{e.g.}
\def\etl{$et ~al.$}
\def\etc{ etc.}

\def\al{\alpha}
\def\bc{\begin{center}}
\def\ec{\end{center}}
\def\eqn{\end{equation}\noindent}
\def\eqnr{\end{eqnarray}\noindent}
\def\beqr{\begin{eqnarray}}

\begin{document}

\catchline{}{}{}{}{} 

\markboth{Awadhesh Prasad}{Existence of  perpetual points in nonlinear dynamical systems and its applications}

\title{Existence of  perpetual points in nonlinear dynamical systems and its applications}

\author{Awadhesh Prasad}

\address{Department of Physics and Astrophysics, University of Delhi, Delhi 110007, India.\\
awadhesh@physics.du.ac.in}

\maketitle

\begin{history}
\received{(to be inserted by publisher)}
\end{history}

\begin{abstract}
A new class of critical points, termed as {\sl perpetual points}, where acceleration becomes zero but
 the velocity remains non-zero, are observed in dynamical systems. 
The velocity at these points is  either maximum or minimum or of inflection behavior.
These points also show the bifurcation behavior as parameters of the system vary.
These perpetual points are useful for locating the  hidden oscillating attractors as well as
co-existing attractors. Results show that these points are important for better understanding of 
transient dynamics in the phase space. The existence of these points confirms whether a system is dissipative  or not.
Various examples are presented, and results are discussed analytically as well as numerically.
\end{abstract}


\section{Introduction}
\noindent 

The stationary  points of  any dynamical system are the ones where
velocity and acceleration  of the system simultaneously become zero. 
These points,
which are important to understand the system in the first step,   are termed as fixed points (FP).
The stability of these fixed points and the motion around  them have been  studied in detail,
and can be found in any standard book on dynamical systems [Ott, 1993; Strogatz, 1994;
Jordan \& Smith, 2009; Kuznetsov, 2004; Kaushal \& Prashar 2008; Lakshmanan \& Rajasekar, 2003;
Arrosmith \& Place, 1990].

In this paper we show that, in nonlinear dynamical systems, there  also exist 
a new class of  points (other than FPs) where acceleration  becomes zero while velocity remains nonzero.
These points can be  termed as {\sl perpetual points} (PP)\footnote{\label{pp} The name  ``perpetual"  is 
used in analogy with
``perpetual motion" which describes a motion that continues indefinitely without any
 external driving force (acceleration) -- see {\sl  http://en.wikipedia.org/wiki/Perpetual{\_}motion}}. 
These points can be traced by considering  the  higher derivative of the velocity vector of the system.
We demonstrate here that  a  study of these  perpetual points is
necessary for  understanding   several new features  of dynamical systems.
In particular,  these points can be useful for locating  the attractors, understanding the dynamics in phase space,
 and confirming whether a system is conservative or not.
To this effect, several examples of different type of dynamical systems are analyzed here:
ranging from   low to high dimension, as well as  from simple to complex ones.
We also observe that these points show the bifurcation behavior as parameters are changed.
 
This paper is organized as follows.
We study  the linear stability around the perpetual points
 in next section. The   distinct features   of PPs and FPs are discussed in  Sec. \ref{diff}.
This is followed by bifurcation analysis  of PPs in Sec. \ref{bif}.
The use of PPs  are discussed in Sec. \ref{appl}. The results are summarized in Sec. \ref{summ}. 

\section{Perpetual points and  stability analysis}
\label{pp-lsa}

Consider a  general dynamical system specified by the equations  

\begin{eqnarray}
\dot{\mathbf{X}}=\mathbf{F}(\mathbf{X},\mathbf{\alpha})
\label{eq:f}
\end{eqnarray}
\noindent
where $\mathbf{X}=(x_1,x_2,...,x_n)^T$ is $n$-dimensional vector of dynamical variables and 
 $\mathbf{F}=(f_1(\mathbf{X}), f_2(\mathbf{X}), ...f_n(\mathbf{X}))^T$ specifies the
 evolution equations (velocity vector) of the system with internal  
parameter $\mathbf{\alpha}$. 
Here $T$ stands for transpose of the vector.
The first order Taylor expansion due to a small perturbation, $\mathbf{\bar{\delta} X}$ \footnote{\label{pert}The separate notations for
perturbations, $\mathbf{\bar{\delta} X}$ and $\mathbf{{\delta} X}$, are used for different Eqs. (\ref{eq:f}) and (\ref{gen}) respectively.} , in Eq. (\ref{eq:f}
)
 around the fixed point $\mathbf{X}_{FP}$ (where $\dot{\mathbf{X}}=0$)\footnote{Taylor expansion of a differentiable  $\mathbf{F}$ can be done at any point in phase space.} leads to 
\begin{eqnarray}
 {\bar{\delta}} \dot{\mathbf{  X}}=\mathbf{F}_{\mathbf{X}^T}(\mathbf{X},\mathbf{\alpha})|_{\mathbf{X}_{FP}} \cdot \mathbf
{\bar{\delta} X} 
\label{eq:df}
\end{eqnarray}
\noindent
where $\mathbf{F}_{\mathbf{X}^T}(\mathbf{X},\mathbf{\alpha})$ is  Jacobian of
size $n\times n$.
The dynamics near    the fixed points depend on the eigenvalues ($\l_i$)
and  the corresponding eigenvectors of this matrix, and the same has  been well 
studied and documented in literature [Ott, 1993; Strogatz, 1994; 
Jordan \& Smith, 2009; Kuznetsov, 2004; Kaushal \& Prashar 2008; Lakshmanan \& Rajasekar, 2003;
Arrowsmith \& Place, 1990].

Acceleration of  the system can be   obtained 
by taking    derivative of Eq. (\ref{eq:f})  with respect to time\footnote{If the velocity 
vector, $\mathbf{F}$, is differentiable.}, viz.
\begin{eqnarray}
\nonumber
\ddot{\mathbf{X}} &=&\mathbf{F}_{\mathbf{X}^T}(\mathbf{X},\mathbf{\alpha})\cdot\mathbf{F}(\mathbf{X},\mathbf{\alpha})\\
\label{gen}
                 &=&\mathbf{G}(\mathbf{X,\alpha})
\label{eq:g}
\end{eqnarray}
\noindent
where $\mathbf{G}=\mathbf{F}_{\mathbf{X}^T}(\mathbf{X},\mathbf{\alpha})\cdot\mathbf{F}(\mathbf{X},\mathbf{\alpha})$
 may be termed as acceleration vector. For convenience,  we drop $\mathbf{X}$
and $\mathbf{\alpha}$ from the arguments of $\mathbf{F}$ and $\mathbf{G}$ from now on.

For  zero acceleration of the system, we set $\ddot{\mathbf{X}}=\mathbf{G}=0$ in Eq. (\ref{eq:g}) which 
gives a set of points where velocity $\dot{\mathbf{X}}$ may be either zero or a nonzero constant.
This set  includes the fixed points ($\mathbf{X}_{FP}$) with zero velocity 
as well as a subset of  new points with nonzero  velocity\footnote{These are possible only in nonlinear systems.}. 
These nonzero velocity  points are the perpetual points ($\mathbf{X}_{PP}$)$^{\ref{pp}}$

The linear stability analysis of Eq. (\ref{eq:g}) around the perpetual points gives
\begin{eqnarray}
\label{eq:dgen}
\delta\ddot{\mathbf{X}} &=& (\mathbf{I}\otimes {\mathbf{\delta {X}}^T})\cdot 
\mathbf{F}_{\mathbf{X}\mathbf{X}^T}\cdot\mathbf{F}+\mathbf{F}_{\mathbf{X}^T}\cdot\mathbf{F}_{\mathbf{X}^T}\cdot\mathbf{\delta X}\\
&=&\mathbf{G}_{\mathbf{X}^T}|_{\mathbf{X}_{PP}}\cdot \mathbf{\delta X} 
\label{eq:dg}
\end{eqnarray}
\noindent 
where $\otimes$ is direct product while $\mathbf{I}$ and $\mathbf{F}_{\mathbf{X}{\mathbf{X}^T}}$ are identity and 
Hessian matrices of
dimension $n\times n$ and $nn\times n$ respectively.
Note that matrix $\mathbf{G}_{\mathbf{X}^T}$
contains both the Jacobian ($\mathbf{F}_{\mathbf{X}^T}$)  and Hessian ($\mathbf{F}_{\mathbf{XX}^T}$) matrices.
For simplicity, the $ij$-element of matrix $\mathbf{G}_{\mathbf{X}^T}|_{\mathbf{X}_{PP}}$ can be written as
\begin{eqnarray}
(\mathbf{G}_{\mathbf{X}^T})_{i,j}=\sum_{k=1}^n[f_k f_{ix_kx_j}+f_{kx_j}f_{ix_k}].
\label{eq:gm}
\end{eqnarray}
\noindent
The general solution of the second order differential equation, Eq. (\ref{eq:dg})
around the perpetual  points for nonzero eigenvalues can be written as
\begin{eqnarray}
\mathbf{\delta X}&=& \sum_i^n  \mathbf{V}_i [c_{i1}\exp{(+\sqrt{\mu_i} t)}
+c_{i2}\exp{(-\sqrt{\mu_i} t)}]
\label{eq:sol}
\end{eqnarray}
\noindent
where $\mu_i$ and $\mathbf{V_i}$ are the  eigenvalues and eigenvectors respectively [Taylor, 2011].  Here
 $c_{i1}$ and $c_{i2}$ are the constants which depend on the initial position $\mathbf{X}(0)$ and 
perturbation $\mathbf{\delta X}(0)$. This solution
 determines the variation of velocity around the perpetual points.
 Note that the solution of equation Eq. (\ref{eq:sol}) and Eq. (\ref{eq:df}) are not the same as acceleration vector$^{\ref{pert}}$,
 Eq. (\ref{eq:g}), may not be invertible to velocity vector, Eq. (\ref{eq:f}).
The  properties of these perpetual  points, using Eqs. (\ref{eq:gm}) and (\ref{eq:sol}),
 are demonstrated  below for some  selected systems.

\subsection{One--dimensional systems:}
\begin{figure}[h]
\begin{center}
\vskip.4cm
\includegraphics [scale=0.6]{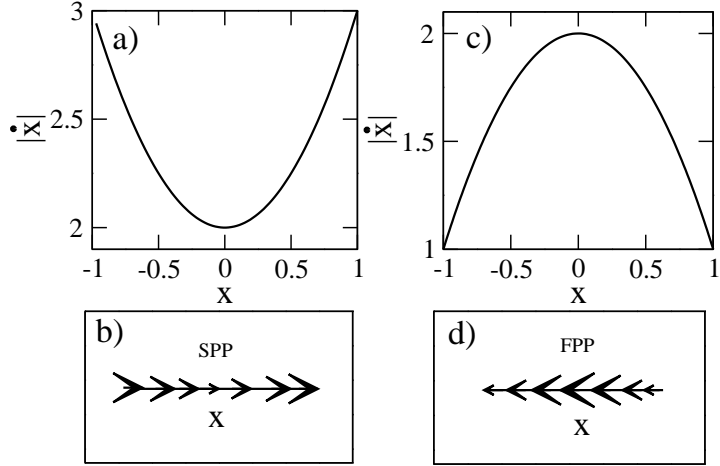}
\end{center}
\caption{Plots in top  row show the magnitude of velocity
while bottom row show the schematic motion in phase-space of Eq. (\ref{eq:1d}).
Left and right panels correspond to the parameter values
 $\alpha=2$ and $\alpha=-2$, respectively.}
\label{fig:1d}
\end{figure}

In order to demonstrate the properties of perpetual points  
we first consider a simple, analytically traceable, dynamical system
\begin{eqnarray}
\dot{ x}&=&x^2+\alpha.
\label{eq:1d}
\end{eqnarray}
For $\al>0$ there is no fixed point and hence system settles at infinity. The
acceleration, $\ddot{x}=2x(x^2+\alpha)$, of this system has  one perpetual point, $x_{PP}=0$.
The velocity at this point is  $\alpha$ while 
 eigenvalue, corresponding to  $\mathbf{G}_{\mathbf{X}^T}|_{\mathbf{X}_{PP}=0}$, is $\mu=2\alpha$.
The general solution of Eq. (\ref{eq:dg}) for  arbitrary  initial 
perturbations $\delta x(0)$ and $\delta \dot{x}(0)$  around this perpetual point is 
\begin{eqnarray}
\delta x&=&c_{11} \exp(\sqrt{2\alpha} t)+c_{12}\exp(-\sqrt{2\alpha}t),
\label{eq:1d221}
\end{eqnarray}
\noindent
where $c_{11}=(\sqrt{2\al} \delta{x}(0)+\delta{\dot{x}}(0))/2\sqrt{2\al}$ 
and $c_{12}=(\sqrt{2\al} \delta{x}(0)-\delta{\dot{x}}(0))/2\sqrt{2\al}$. 
Here, $\delta{\dot{x}}(0)=2{x}(0)\delta{{x}}(0)$
from Eq. (\ref{eq:1d}).
With  simple manipulation of this solution  and its derivative, we  get the relation
\begin{eqnarray}
2\al \delta {x}^2-\delta \dot{x}^2&=&[2\al-4x(0)^2]\delta{x}(0)^2, 
\label{eq:1d1s}
\end{eqnarray}
which clearly shows that both
 $\delta x$ and $\delta \dot{x}$  decrease simultaneously as a trajectory approaches the PP
 \ie, in the neighborhood of  this type of perpetual point where $\mu>0$  system moves  slowly.
Therefore, this point is termed as slow perpetual point (SPP)\footnote{\label{fast-slow} The nomenclature  ``fast" or  ``slow" for  perpetual points is used because   
system moves with ``maximum" or ``minimum" velocity at these
points.  This  nomenclature is considered in the spirit of definition  ``stationary point"
 where velocity is zero.}.
The magnitude of  velocity, $|\dot{x}|$, is shown in Fig. \ref{fig:1d}(a) while direction of motion in phase-space
is indicated in (b).
Here the size of the arrows indicate the magnitude of velocity. 
Note that the trajectory passes through the PP, in contrast to the FP where
 trajectories never cross.

However for $\al<0$, Eq. (\ref{eq:1d}) has  two fixed points, $x_{FP}=-\sqrt{|\al|}$ and $\sqrt{|\al|}$,
which are respectively stable and unstable \ie, initial conditions $x(0)<\sqrt{|\al|}$ lead to 
$x_{FP}=-\sqrt{|\al|}$ otherwise settle at  infinity.
The perpetual point for this case is 
$x_{PP}=0$ where velocity and eigenvalue are  $-|\alpha|$ and $\mu=-2|\alpha|$ respectively.
The general solution of Eq. (\ref{eq:dg})   around this  PP  is
\begin{eqnarray}
 \delta x=c_{11}\cos( \sqrt{2|\al|} t)+c_{12} \sin(\sqrt{2|\al|} t) 
\label{eq:1d22s}
\end{eqnarray}
\noindent

where $c_{11}=\delta{x}(0)$ and $c_{12}=\delta \dot{x}(0)/\sqrt{2|\al|}=2x(0)\delta x(0)/\sqrt{2|\al|}$, 
which along with its derivatives give the relation
\begin{eqnarray}
2{\lvert\al\rvert } \delta {x}^2+\delta \dot{x}^2&=&[2|\al|+4x(0)^2]\delta{x}(0)^2. 
\label{eq:1d2s}
\end{eqnarray}
This indicates that as $\delta x$ decreases \ie, trajectory reaches to $x_{PP}=0$,
  $\delta \dot{x}$ increases as shown in Figs. \ref{fig:1d}(c) and \ref{fig:1d}(d).
 This suggests that as the system reaches  PP, its  velocity becomes higher and higher, and hence it moves
 faster. Therefore this type of PP, where $\mu<0$, is termed as faster perpetual point (FPP)$^{\ref{fast-slow}}$.

 For special cases when  eigenvalues are zero, Eq. (\ref{eq:sol}) can be written as
 $\mathbf{\delta X}=  \mathbf{V}_i [c_{i1} +c_{i2} t]$ where
$\mathbf{V}_i$ is eigenvector corresponding to the zero eigenvalue. For an example,
consider system $\dot{x}=(x-1)^3-\alpha, \alpha>0$ which  has one unstable fixed point,
 $x_{FP}=1+\alpha^{1/3}$, with $\lambda=3\alpha^{2/3}$ and one perpetual point, $x_{PP}=1$, with  $\mu=0$. 
Since $\mu=0$ therefore there is no extremum in velocity. However the trajectory starting from 
$1<x(0)<2$ approaches  $x=1$ with high velocity and then becomes constant at PP. Once it passes through PP,
velocity increases again. This can be seen as an inflection point in velocity-distance plot, or also
from the solution $\delta x=\delta x(0)+t\delta \dot{x}$.

We observed similar results for other one-dimensional systems also -- see Eq. (\ref{eq:bif}).
The formalism, (Eq. (5)),  which is  verified here for  1D-systems,
now can be  extended  to  higher dimensional systems, as given below.

\subsection{Two--dimensional systems:}
\begin{figure}
\begin{center}
\includegraphics [scale=0.5] {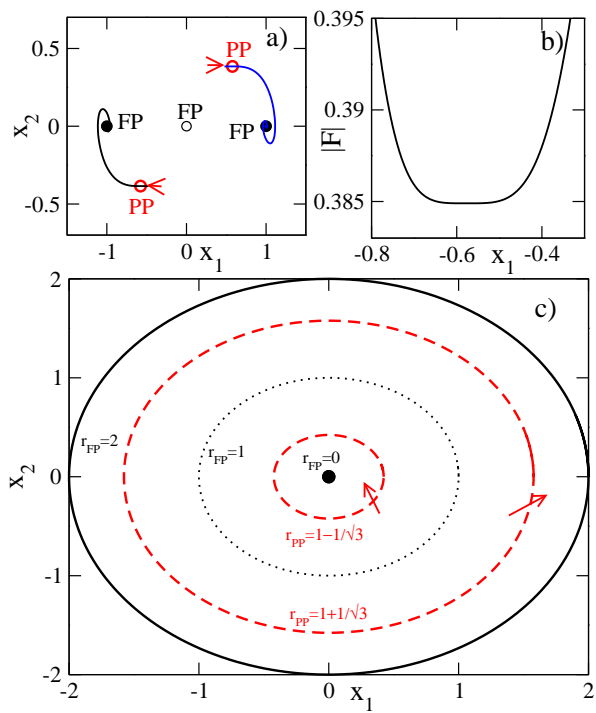}
\end{center}
\caption{The (a) dynamical behavior around the  fixed and perpetual points,
(b) the magnitude of  velocity as a function of $x_1$ along
 the vector $(1,0)^T$ near  one of the PPs of Eq. (\ref{eq:2da}) for $\alpha=1$ and $\beta=1$.
(c) The orbits and loci of fixed and perpetual points respectively of Eq. (\ref{eq:2db})-see
 text for details.  Arrows show the directions of flow.} 
\label{fig:2d}
\end{figure}

Here we   consider the  celebrated Duffing oscillator [Ott, 1993]
\begin{eqnarray}
 \ddot{x}+\beta \dot{x}-\alpha x+x^3=0. 
\label{eq:2da}
\end{eqnarray}
\noindent
It has three fixed points at $x_{FP}= 0$ and $\pm \sqrt{\alpha}$ for $\alpha>0$. The first  point is unstable 
while the last two are stable, which are shown in Fig. \ref{fig:2d}(a) in ($x_1=x, x_2=\dot{x}$) plane with
open and filled circles. 
The perpetual points of this system are
$(-\sqrt{\al/3}, -\frac{2\al}{3\beta}\sqrt{\al/3})$ and $(\sqrt{\al/3},
 \frac{2\al}{3\beta}\sqrt{\al/3})$\footnote{Here  PPs are in fact
the   inflection points of the potential $x^4/4-\alpha x^2/2$,  Eq. (\ref{eq:2da}).}. The stability analysis  suggests that
 eigenvalues for both  the  PPs are  $\mu=(\beta^2\pm\sqrt{\beta^4+16\alpha^2/3})/2$.
Since the largest eigenvalue  is positive for these PPs   the velocities
should be  slower. This is confirmed in Fig. \ref{fig:2d}(b) where the magnitude of  velocity,
 $|F|=\sqrt{\dot{x}_1{^2}+\dot{x}_2^2}$, along the velocity vector, $(1,0)^T$ at one
PP is  plotted. This clearly shows that the velocity is minimum at this PP.

In another example of two-dimensional system we consider 
a system having  coexisting attractors, a stable fixed point and a limit cycle, 
\begin{eqnarray}
\nonumber
\dot{r}&=&r(r-1)(2-r)\\
\dot{\theta}&=&\omega.
\label{eq:2db}
\end{eqnarray}
\noindent
Here $r=\sqrt{x_1^2+x_2^2}$ and $\theta=\tan^{-1}{x_2/x_1}$ correspond to the radius
 and the phase of the system with frequency $\omega$. 
The fixed point analysis of this system gives  two coexisting attractors at 
$r=0$ (stable fixed point) and $r=2$ (limit cycle), and  one unstable periodic 
orbit (UPO) at $r=1$.  These, stable and unstable features,  are shown in Fig. \ref{fig:2d}(c) with  
solid  and dotted  lines respectively. The  perpetual points of this system are,  $r_{PP}=1\pm1/\sqrt{3}$ 
which form  closed loci,  shown in Fig. \ref{fig:2d}(c) 
with red-dashed lines. The eigenvalues at perpetual points are
$\mu=-6r_{PP}(1-r_{PP})^2(2-r_{PP})$. Since $r_{PP}<2$, therefore $\mu<0$,  which implies  
that velocities have maxima at these points. The other two-dimensional systems are also considered in Sec. \ref{appl-cons}.

\vskip.2cm
\noindent
\subsection{Three--dimensional systems:}
\label{3d}
\begin{figure}
\begin{center}
\includegraphics [scale=0.5]{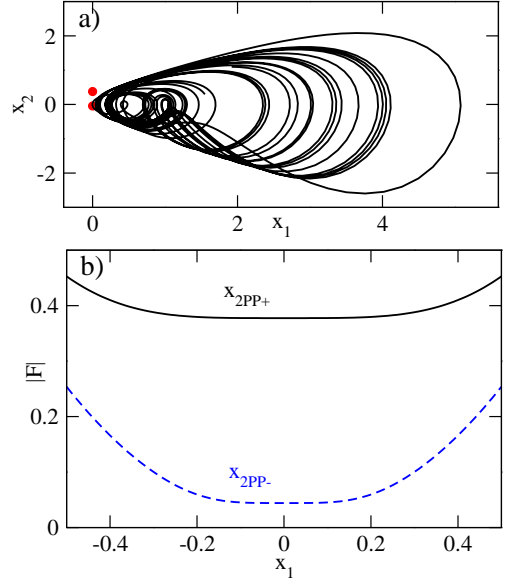}
\end{center}
\caption{ (a) A hidden chaotic attractor and 
 (b) the magnitude of  velocities, along the velocity vectors passing through 
 the perpetual points.}
\label{fig:3d}
\end{figure}

The  self-excited attractors, whose  basins  intersect with the neighborhood of
the fixed points (\eg~ Lorenz, Rossler, Chua, \etc)  have been studied in
detail [Ott, 1993; Strogatz, 1994;
Jordan \& Smith, 2009; Kuznetsov, 2004; Kaushal \& Prashar 2008; Lakshmanan \& Rajasekar, 2003;
Arrowsmith \& Place, 1990]. Very recently, a new type of attractors called  
hidden attractors, that don't intersect with the neighborhood of  any FP
have been reported [Kuznetsov \etl~ 2010; Leonov \etl~ 2011a; Leonov \etl~ 2011b;
 Leonov \etl~ 2013].
Due to the absence of unstable FP in its
neighborhood these type of attractors are less tractable. 
Therefore, it is  also difficult to understand their characteristic 
behavior [Kuznetsov \etl~ 2010; Kuznetsov \etl~ 2011; 
 Leonov \etl~ 2010; Leonov and Kuznetsov 2011;
Leonov \etl~ 2011a, Leonov \etl~ 2011b;   Leonov \etl~ 2011c;  Chaudhuri \& Prasad 2014]. 
  As an example, we  consider such a system that has no fixed point
 and  provides  hidden attractor [Wang \&  Chen, 2013], 
\begin{eqnarray}
\nonumber
\dot{x}_1&=&x_2 \\
\nonumber
\dot{x}_2&=&x_3 \\
\dot{x}_3&=&-x_2+3x_2^2-x_1^2-x_1x_3+\alpha.
\label{eq:3d}
\end{eqnarray} 
\noindent
This system doesn't have any fixed point for parameter $\alpha<0$.
A typical chaotic hidden attractor is shown in Fig. \ref{fig:3d}(a)
for $\alpha=-0.05$.  Because there is no fixed point in this system 
at this value of parameter,  this attractor is termed as  hidden one.

The perpetual points for this systems are $(0, x_{2PP\pm},0)$ where $x_{2PP\pm}
=(1\pm\sqrt{1-12\alpha})/6$. These  are shown  in
Fig. \ref{fig:3d}(a) with  red-circles. The 
largest eigenvalues $\mu_i$, corresponding to the matrix $\mathbf{G}_{\mathbf{X}^T}$,   at these PPs 
are $6x_{2PP\pm}-1$ and $[5x_{2PP\pm}-1\pm\sqrt{13x_{2PP\pm}^2-14x_{2PP\pm}-4\alpha+1}]/2$
which are  $+$ve  for the both points\footnote{The eigenvalues, $\mu_i$ at $x_{2PP+}$ and $x_{2PP-}$ are
($1.2649, 0.4437\pm i 0.7470$)  and ($0.0684,-1.2649, -1.2892$) respectively.}.  Therefore the velocities
at these points should be slower, which is confirmed in Fig. \ref{fig:3d}(b) where 
the magnitude of velocities are plotted along the velocity vectors at these points\footnote{\label{para} The velocity vectors  at PP $(0,x_{2PP\pm},0)$ are
 $(x_{2PP\pm},0,-x_{2PP\pm}+x_{2PP\pm}^2-\alpha)^T$. 
The velocities are calculated along  the lines which are estimated using
the   parametric ($p$) form of the equation of a
 line passing through these PPs:
 $x_1=px_{2PP\pm}$, $x_2= x_{2PP\pm}$ and $x_3=p(-x_{2PP\pm}+x_{2PP\pm}^2+\alpha)$ which
give as relation $x_3=x_1 (-x_{2PP\pm}+x_{2PP\pm}^2+\alpha)/x_{2PP\pm} $.}.
An another system with no fixed point is also considered  in Sec. \ref{appl-hid}.

\section{Perpetual Points vs. Fixed Points}
\label{diff}

\begin{center}
\begin{table}
\tbl{Comparative properties of FP and PP. Here, $\mu_m$ is the largest nonzero
eigenvalue among $\mu_i$.}
{    \begin{tabular}{| c | c | c |}
    \hline
~&Fixed Point& Perpetual Point\\
\hline
Velocity, $|F|$ & $|F|=0$ &$|F|\ne0 $ \\
~&~ & Extremum if $\mu_m \lessgtr 0$\\
~&~ & Inflection if  $\mu_i=0~~ \forall i$ \\
\hline
Acceleration & $0$ & $0$\\
\hline
$\l_i$  & $+, -, 0$ & At least one of $\l_i=0 $\\
$ $ [Eq. (\ref{eq:df})] &  & \\
\hline
$\mu_i$  & $\mu_i=\lambda_i^2$  & $ +, -, 0$\\
$ $ [Eq. (\ref{eq:dg})] &  & $ $\\
\hline
Motion  & Stable or unstable & Closed locus\\
 of points & periodic orbits&   \\
\hline
Trajectory & Moves towards  & Passes through \\
~ & or Moves away  & ~ \\
\hline
  \end{tabular}}
\end{table}
\end{center}

In this section we present the  comparative properties of PPs and FPs, as  summarized in Table I: ({\sl i})
The  velocity at the perpetual points  is  either maximum or minimum or of inflection behavior
while zero for fixed points. However, as per definition, acceleration is zero at both  the PPs and FPs. 
({\sl ii}) The eigen values, $\lambda_i$, corresponding to Eq. (2) are either negative or positive or zero at
 fixed points. However  one of $\l_i$ must be zero at PPs. Its reason can be explained as follows. 
In general, if $\mathbf{F}_{\mathbf{X}^T}\cdot\mathbf{F}=0 $ then there is either
a trivial solution $\mathbf{F}=0$ or infinite number of  solutions. Since  $\mathbf{F}\ne 0$ at the perpetual points
 therefore a finite number of solutions exists if and only if $\mathbf{F}_{\mathbf{X}^T}$ admits at least
one  zero eigenvalue corresponding to the eigenvector  $\mathbf{F}$.
({\sl iii}) Here, as we have observed in previous section,  the value of $\mu_i$ could be either positive or negative or zero at PP. 
 At FPs, first term in Eqs. (\ref{eq:g}) becomes zero due to $\mathbf{F=0}$,
and hence $\mu_i$ is related to $\l_i$  as $\mu_i=\l_i^2$. 
({\sl iv }) A trajectory   passes through a PP whereas  it either moves  away or settles on a fixed point.
Note that  in some cases, for example Eq. (\ref{eq:2db}), PP forms a closed
 locus as shown in Fig. \ref{fig:2d}(c) whereas  FP forms either stable or unstable periodic orbits.

\section{Bifurcations}
\label{bif}
\begin{figure}
\begin{center}
\includegraphics [scale=0.55]{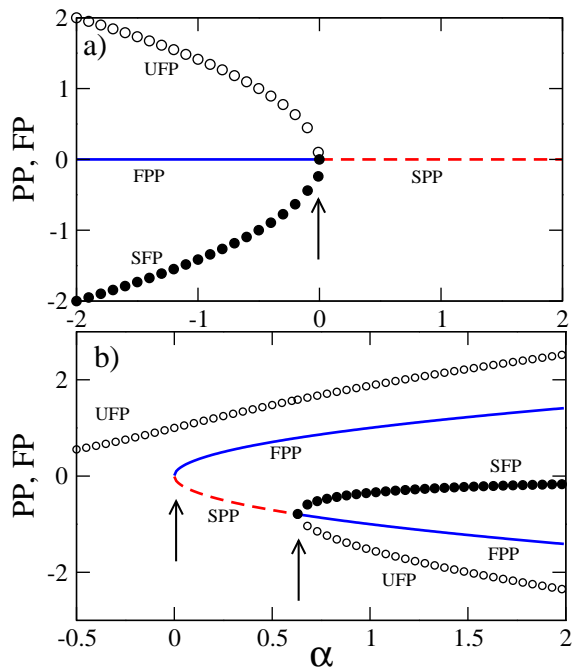}
\end{center}
\caption{ The bifurcation diagrams of systems (a) Eq. (\ref{eq:1d}) and (b) Eq. (\ref{eq:bif}) as a function of
 parameter $\alpha$. Arrows show the bifurcation points. Filled and open circles
are the stable and unstable FPs while Solid-blue and dashed-red lines are FPP and SPP respectively.}
 \label{fig:bif}
\end{figure}

In order to explore the possibilities of
occurrence of bifurcations of perpetual points as a function of parameters, say $\alpha$,
we first consider system Eq. (\ref{eq:1d}). Shown in  Fig. \ref{fig:bif}(a)  are the filled and open circles for
stable and unstable fixed points along with the solid-blue and dashed-red line for fast and slow PPs respectively.
This shows that below $\alpha=0$ there are two fixed points (stable and unstable) and one
 fast perpetual point at $x=0$\footnote{\label{bifnote}In one-dimensional systems there must be at least 
one fast perpetual point in between two fixed  points.}. 
As parameter $\alpha$ passes through  $\alpha=0$  there is no fixed point but
the  fast perpetual point becomes slower one \ie~ there is change in magnitude  of velocity from maximum to minimum 
 across $\alpha=0$.  Therefore this  change in  dynamics  may be termed as bifurcation of PP ($x_{PP}=0$). 

In another example, we construct a system,
\begin{eqnarray}
\dot{x}&=&x^3-3\alpha x-1.
\label{eq:bif}
\end{eqnarray}
 \noindent
Depending on the parameter, $\alpha$, this system has either one or three  fixed points  which are
 shown in bifurcation diagram, Fig. \ref{fig:bif}(b). The  open  and filled circles
correspond to the stable and unstable fixed points (denoted as SFP and UFP respectively).
This system has two perpetual points,  $x_{PP}=\pm\sqrt{\alpha}$, which are
 shown in Fig. \ref{fig:bif}(b)  with solid and dashed
lines. At $\alpha=0$, the fast ($\mu=6(-2\alpha^2-\sqrt{\alpha})$) and 
the slow ($\mu=6(-2\alpha^2+\sqrt{\alpha})$) perpetual points 
(shown as solid and dashed lines respectively) are created.
 As parameter $\alpha$ is increased 
further,  slow perpetual point becomes fast at $\alpha\sim 0.62$ where
saddle-node bifurcation$^{\ref{bifnote}}$ for fixed points occurs. 
 We observe similar bifurcations in  other systems (from Ref. [Strogatz, 1994]) also, 
which confirm that the perpetual points do show  bifurcations behavior in parameters space.

\section{Applications}
\label{appl}

\subsection{Conservative vs. Dissipative}
\label{appl-cons}

Checking a system whether it is  conservative or not is an important problem in nonlinear dynamics.
 There are various ways to confirm whether a system is conservative or 
not [Strogatz, 1994; Ott, 1993]. In this work we show that this can be checked
in a simple way by searching the perpetual points.
Let us consider an autonomous Hamiltonian system
 $H=H(q,p)$ of one-degree of freedom (the following results can be extended for 
the systems with $n-$degrees of freedom  also).
Here $q$ and $p$ are the generalized co-ordinate and the momentum respectively.
The  equations of motion are

\begin{eqnarray}
\nonumber
\dot{q}&=&\frac{\partial H}{\partial p}, \\
\dot{p}&=&-\frac{\partial H}{\partial q}.
\label{eq:h1}
\end{eqnarray}

\noindent
The next  higher order derivatives of these equations are given as

\begin{eqnarray}
\nonumber
\ddot{q}&=& \frac{\partial^2 H}{\partial p \partial q }\dot{q}+
 \frac{\partial^2 H}{\partial p\partial p }\dot{p} \\
\ddot{p}&=&- \frac{\partial^2 H}{\partial q \partial q }\dot{q}-
\frac{\partial^2 H}{\partial p\partial q }\dot{p},
\label{eq:h2}
\end{eqnarray}

\noindent
which can be recast in the matrix form as
\begin{eqnarray}
\nonumber
& &\\
\begin{bmatrix}
\ddot{q}\\
\ddot{p}
\end{bmatrix}
&=&
\begin{bmatrix}
\frac{\partial^2 H}{\partial p \partial q } & \frac{\partial^2 H}{\partial p^2 }\\
-\frac{\partial^2 H}{\partial q^2} &-\frac{\partial^2 H}{\partial p \partial q}
\end{bmatrix}
\begin{bmatrix}
\dot{q}\\
\dot{p}
\end{bmatrix}\\
\Rrightarrow \ddot{\mathbf{X}}&=& \mathbf{F}_{\mathbf{X}^T}\cdot\mathbf{F} \mbox{~~~~~[cf. Eq. (\ref{gen})]}
\end{eqnarray}

\noindent
where $\mathbf{X}=(q,p)^T$.
For getting perpetual points we must have $\ddot{\mathbf{X}}=0$. However, 
if $\mathbf{F}_{\mathbf{X}^T}\cdot\mathbf{F}=0 $ then there is either
the trivial solution $\mathbf{F}=0$ or there are infinite number of  solutions. 
Since  $\mathbf{F}\ne 0$ at the perpetual points,
the finite solution exists if and only if $\mathbf{F}_{\mathbf{X}^T}$ admits at least
one  zero eigenvalue corresponding to the eigenvector  $\mathbf{F}$.
However, due to the special structure of the Hamiltonian, the trace of $\mathbf{F}_{\mathbf{X}^T}$ is zero. 
This means that the determinant of $\mathbf{F}_{\mathbf{X}^T}$ should  also be zero. 
But $\mathbf{F}_{\mathbf{X}^T}=0$ is possible
 only at the fixed point,  hence the  existence of  PP is ruled out. This suggests that the existence of
PP can confirm   dissipation in the systems.  This is demonstrated in the following examples.

\begin{figure}
\begin{center}
\includegraphics [scale=0.95]{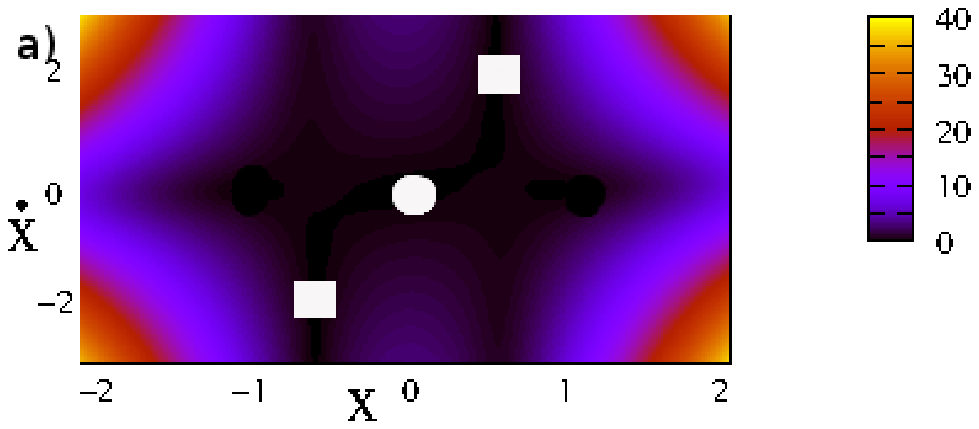}\\
\includegraphics [scale=0.95]{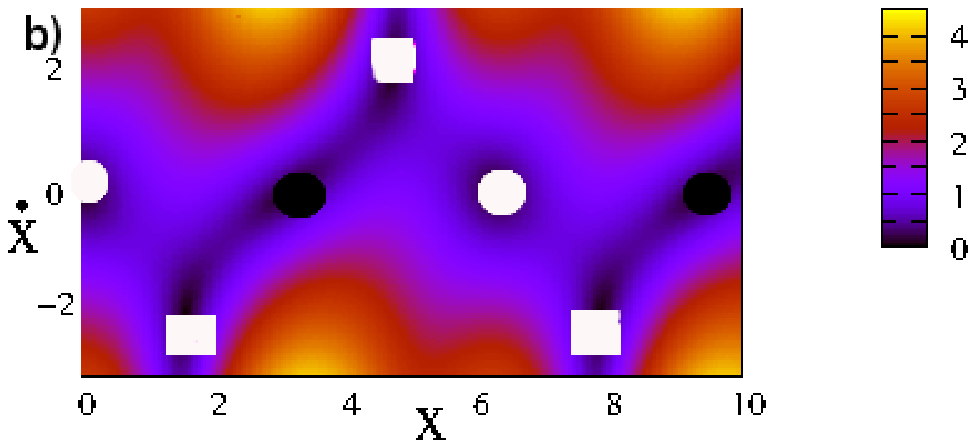}
\end{center}
\caption{Acceleration in $x-\dot{x}$ plane for (a) Duffing system, Eq. (\ref{eq:2da}), and (b) plane
 pendulum, Eq. (\ref{eq:pen}), at parameters $\alpha=1~\&~\beta=3$ and  $\alpha=2~\&~\beta=1$ respectively.}
 \label{fig:cons}
\end{figure}

\vskip.2cm
\noindent
{\sl Duffing systems:} We first consider the  Duffing system, Eq. (\ref{eq:2da}), which has two  
 perpetual points
$(-\sqrt{\al/3}, -\frac{2\al}{3\beta}\sqrt{\al/3})$ and $(\sqrt{\al/3},
 \frac{2\al}{3\beta}\sqrt{\al/3})$.
These points are shown in Fig. \ref{fig:cons}(a) with squares. This clearly indicates  that for zero   strength of 
dissipation, $\beta \to 0$,  the PP goes to infinity \ie~  there is no PP in the conservative limit,  
and hence at $\beta \to 0$ system is conservative.

\vskip.2cm
\noindent
{\sl Plane pendulum}: Consider the  equation of  a plane pendulum 
\begin{eqnarray}
\ddot{q}+\beta\dot{q}+\alpha \sin{q}&=&0
\label{eq:pen}
\end{eqnarray}
\noindent
where $\alpha$ and $\beta$ are the system parameters. For $\beta=0$, system
is conservative and Hamiltonian (total energy) is constant of motion.  Its  equations of motion can be written 
 as
\begin{eqnarray}
\nonumber
\dot{q}&=& p\\
\dot{p}&=&\beta p -\alpha \sin{q}.
\end{eqnarray}

\noindent
The fixed points of this system are $(\pm k\pi, 0)$ where $k=\pm1, \pm2 ...$ 
The perpetual points for this system are $(\pm (n+1/2) \pi, \pm \alpha/\beta$). These points, FPs (circles) and
PPs (squares) are shown in Fig. \ref{fig:cons}(b). The color represents the acceleration
 $\sqrt{\ddot{q}^2+\ddot{p}^2}$. 
Similar to the Duffing system, it clearly indicates that as strength of dissipation, $\beta \to 0$ 
 the PP goes to infinity \ie~
there is no PP in the conservative limit. Here also  in strong dissipation, \ie~ $\beta\to\infty$,  ${\dot{p}}_{_{\tiny PP}}\to 0$.
 Note that these PPs are at the inflection  point of the potential of the Hamiltonian.

\vskip.2cm
\noindent
{\sl Lotka-Volterra system:} We consider another celebrated system which is not in the Hamiltonian form.
 The governing equations of  this system [Lotka, 1920] are 
\begin{eqnarray}
\nonumber
\dot{x}&=& x(a-by)\\
\dot{y}&=&y(cx-d)
\label{eq:lv}
\end{eqnarray}

\noindent
where $x$ and $y$ are prey and predator populations respectively while $a,b,c,$ and $d$ are 
the system's parameters. Note that this system is quite different in structure   as compared to the
Duffing and Pendulum systems. However, after certain transformation [Lotka, 1920] it can be shown
 that this system has one invariant $I=d\log{x}-cx+a\log{y}-by$,  suggesting that the system is conservative.
For checking the existence  of PP in this system, we examine the acceleration vector
$$
\begin{bmatrix}
G(X)
\end{bmatrix}
=
\begin{bmatrix}
x(a-by)^2-bxy(cx-d)\\
cxy(a-by)+y(cx-d)^2
\end{bmatrix}.
$$
\noindent
The fact is that  $G(X)\ne0$ for positive parameters. This implies that  there are no perpetual points,
 and hence it is a conservative system.
All these examples suggest that, in a system,  if there   exists a PP   then it is 
 dissipative otherwise  conservative.

\subsection{Locating hidden and coexisting attractors}
\label{appl-hid}
The understanding of hidden attractors (cf. Sec. \ref{3d}), as compared to the excitable ones, is
difficult due to the absence of fixed points.  Even to locate the  hidden attractors 
in a given system, one requires proper search method consisting of analytical and numerical
techniques [Leonov \etl~ 2010; Leonov \etl~ 2011b; Bragin \etl~ 2011]. 
In this work we show  that  we can use perpetual points to locate such hidden attractors.

\begin{figure}
\begin{center}
\includegraphics [scale=0.55]{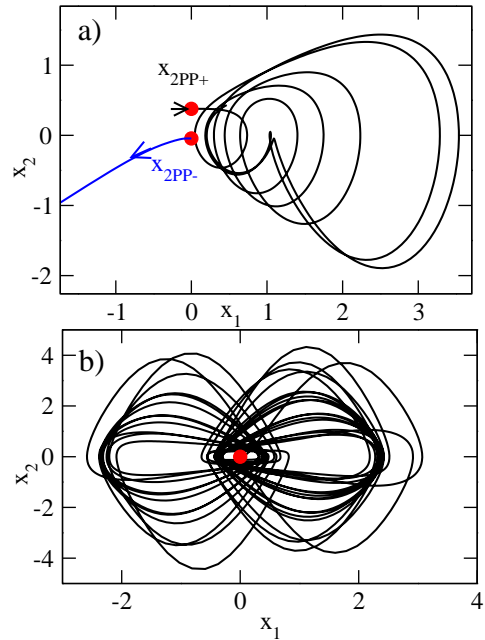}
\end{center}
\caption{The transient trajectories starting from perpetual points (red circles) for systems
(a) Eq. (\ref{eq:3d}) and (b) Eq. (\ref{eq:nh}).}
 \label{fig:hd}
\end{figure}

Consider the system, described by Eq. (\ref{eq:3d}),  which has no fixed point but has one chaotic hidden attractor as shown in 
 Fig. \ref{fig:3d}(a) (see Sec. 2).  It has two  perpetual points,
 $(0,x_{2PP+},0)$ and  $(0,x_{2PP-},0)$ as shown in Fig. \ref{fig:hd}(a) with red-circles. 
The transient trajectories starting from these PPs  are shown in Fig. \ref{fig:hd}(a). It clearly shows that
the trajectories starting from the former one goes to the  hidden attractor (Fig. \ref{fig:3d}(a))
where as the  later one goes to infinity [Chaudhuri \& Prasad 2014].
This confirms that the perpetual point,  $(0,x_{2PP+},0)$, is useful to locate the 
hidden attractor.

We consider  another example, the Nose-Hoover system  [Jafari \etl~ 2013]
\begin{eqnarray}
\nonumber
\dot{x}_1&=&x_2\\
\nonumber
 \dot{x}_2&=&-x_1-x_2x_3\\
\dot{x}_3&=&x_2^2-\alpha.
\label{eq:nh}
\end{eqnarray}

\noindent
This system doesn't have any FP but shows  chaotic motion ($\alpha=1$), as shown in Fig. \ref{fig:hd} (b). 
This system has one PP which is $(0,0,0)$. The trajectory starting from 
the perpetual point  leads to the chaotic attractor. 
As in  other similar systems (from Ref. [Jafari \etl~ 2013]) without FP,  but having oscillations can 
also be located. These examples   confirm that  hidden attractor can be located with PP.
These numerical results  also indicate that hidden attractors (oscillating ones) also have 
 reference point similar to the fixed points of self-excited oscillations. However, a detailed
connection between the perpetual points and the hidden attractors needs to be 
established. Perhaps there is a need to develop a  manifold type of theory  for perpetual points as exists for
 fixed points [Kuznetsov 2004].

We  observed that the coexisting attractors can  also be located using perpetual points. We can easily see
in Fig. \ref{fig:2d}(a) that the trajectories starting from different perpetual points
go to different  stable FPs, either $(0,-\sqrt{\al})$  or $(0,\sqrt{\al})$.
  The trajectories passing through the perpetual points 
 are shown in Fig. \ref{fig:2d}(a) by black and blue lines. 
The conclusion that the PPs can be used to locate the coexisting  attractors
can also be drawn form Fig. \ref{fig:2d}(c) for limit cycle cases where initial 
conditions starting from the perpetual points $r_{PP}=1-1/\sqrt{3}$ and $r_{PP}=1+1/\sqrt{3}$ lead to the
stable fixed point ($r_{FP}=0$) and stable limit cycle ($r_{FP}=2$) respectively.
These examples suggest that  the knowledge of  perpetual points is useful for locating   the  hidden and coexisting 
attractors\footnote{The multi-stable 
attractors [Pisarchik, \& Feudel, 2014] where number of attractors are very large \eg~ in conservative 
limit, are yet to be studied.}.

\subsection{Transients} 
In phase space the velocity or the acceleration of a system doesn't remain constant 
(except for circular motion). 
The accelerations in the  $x-\dot{x}$ plane for the systems described in Eqs. (\ref{eq:2da}) and (\ref{eq:pen})
are shown in Fig. \ref{fig:cons} (color). 
The accelerations are different in different regions
of phase space. Particularly at the PPs, 
accelerations are zero while velocities are extremum (Fig. \ref{fig:cons}). 
Therefore, as a  trajectory moves in phase space
the durations of travel for the equal distances are different. The variation of velocity also  effects the transient
time\footnote{Transient time is defined as the duration in which a system, starting from an initial condition,
reaches to the asymptotic attractor.} \eg~ If we start the system from an initial condition
for which the trajectory passes through the neighborhood of  perpetual points, where velocity is extremum (Fig. \ref{fig:cons}),
then it requires either longer or shorter time (corresponding to SPP or FPP respectively)
 to reach the attractor. 
Therefore,  in many applications where understanding of transients is  important
[Hastings, 2001; Ovaskainen \& Hanski, 2002], location of PPs  in phase space (Fig. \ref{fig:cons})
 is useful.

\section{Summary}
\label{summ}

{\sl Summary:} A new set of  points, termed as {\sl perpetual points} where  
velocity remains nonzero while acceleration becomes zero, are observed.
At  these points the velocity can be  either maximum or minimum or of inflection behavior.
The  properties of perpetual points, along with the fixed points, are
summarized in Table I (Sec. \ref{diff}).

The perpetual points, each having its extremum value of velocity, influence the transient dynamics.
Therefore, the study of PPs is essential to understand   many
natural phenomena, \eg,  experiments where transients 
persist for   longer time, neural processing [Hastings, 2001; Ovaskainen \& Hanski, 2002], 
transient chaos, and in ecology  where the transient dynamics play an important role. 
Using these perpetual points one can also locate
the presence of co-exiting  as well as hidden attractors. 
The existence of PPs in a system  can be used to confirm whether the system is conservative or not.

It  is easier to examine the magnitude of the velocities in lower dimensions 
(particularly one and two dimensional systems). 
In case of  high dimensional systems  it is  difficult to visualize  the extrema of the velocities in phase space.  
Here, the  presented generalized formulation (cf. Eqs. (\ref{eq:dgen}) and (\ref{eq:gm})) can  easily be applied
to any (lower or higher) dimensional systems.
This work also shows that it may be useful to consider
the higher derivatives  (without any additional initial conditions) of the velocity vector for
better understanding of  nonlinear dynamical systems\footnote{The  higher-order derivatives of a nonlinear system
 may give either FPs/PPs only or an entirely  new set of points.
The physical meaning of the higher derivatives and the  corresponding  new points need 
to be explored.}.
Here  we have demonstrated  the analysis  for several different type of systems, which
suggest that the presented results  are quite general.


\nonumsection{Acknowledgments} \noindent 
Author  thanks B. Biswal, R. S. Kaushal, M. D. Shrimali and R. Ramaswamy for critical comments on this manuscript.
  The financial supports from the DST, Govt. of India  and DU-DST PURSE  are also acknowledged.

\end{document}